\documentclass[12pt]{article} %{slides}

\usepackage{amsmath,amssymb,amsfonts,pstricks,pst-plot,pst-node,a4wide}
\usepackage{layout,array}
\usepackage[T1]{fontenc}
\newtheorem{theorem}{Theorem}
\def\bar{\overline}
\def\Hl{\xi_\ell}
\def\Hn{\xi_n}
\def\Hm{\xi_m}
\def\Hmb{\xi_{\bar m}}
\def\mb{\bar m}
\def\o{{o}}
\def\i{{\iota}}
\def\sfdual{{}^{\raise2pt\hbox{$\scriptstyle{-}$}}}

\def\GHPL{\operatorname{\mathcal L}}
\def\thorn{\operatorname{\text{\th}}}
\def\eth{\operatorname{\text{\dh}}}

\begin{document}
\begin{center} % TITLE
{\large\bf The general homothetic equations}
\end{center}

\begin{narrower}

\medskip
\noindent {\bf John D.~Steele}

\smallskip
\noindent {\small School of Mathematics, University of New South Wales,
Sydney, NSW 2052, Australia.

\smallskip
\noindent  email: j.steele@unsw{.}edu{.}au}

\medskip
\noindent {\bf Abstract} 

\noindent 
In an earlier paper [6] the author wrote the homothetic equations for
vacuum
solutions in a first order formalism allowing for arbitrary alignment of the
dyad. This paper generalises that method to homothetic equations in non-vacuum
spaces and also provides useful second integrability conditions. An application
to the well-known Petrov type O pure radiation solutions is given.

\medskip
\noindent PACS number 0420J


\end{narrower}

\section{Recap}

In [6] I gave the homothetic Killing equations in vacuum, written out in a
first order form without the assumption that the spinor dyad used was aligned to
either the symmetry or the curvature in any way, and indeed allowing for the
dyad to be
non-normalised.
In this paper I will generalise to non-vacuum. Conventions and notation will follow
Penrose and Rindler [7].

A homothetic vector $\xi^a$ by definition satisfies the equation
\begin{equation}\xi_{a;b}= F_{ab} + \psi g_{ab}\label{eq:Killing}.\end{equation}
Here $\psi$, the {\bf divergence}, is a constant, and $F_{ab}$ will be called 
the {\bf homothetic bivector}.

Let $\{\o^A,\i^A\}$ be a spinor dyad, with
$\o_A\i^A=\chi$. A complex null tetrad is related to this dyad in the standard 
way:
$$\ell^a=\o^A\o^{A'};\quad n^a=\i^A\i^{A'};\quad m^a=\o^A\i^{A'};\quad
\mb^a=\i^A\o^{A'},$$
([7], (4.5.19)), and $\ell_an^a = -m_a\mb^a = \chi\bar\chi$.
As in [6], we define components of the homothety:
\begin{equation}
\xi_a =\Hn\ell_a + \Hl n_a - \Hmb m_a - \Hm \mb_a,
\end{equation} 
with $\{\ell^a,n^a,m^a,\mb^a\}$ a
Newman-Penrose tetrad. Thus, for example, $\chi\bar\chi\Hl=\xi_a\ell^a$.

For the homothetic bivector
$F_{ab}$ we define its anti-self dual by
\begin{align}
\sfdual{F}_{ab}& =\frac{1}{2}\left(F_{ab}+i{}F^*_{ab}\right)
\end{align}
and then
\begin{equation}
\sfdual{F}_{ab} =(\chi\bar\chi)^{-1}\left( 2\phi_{00}\,\ell_{[a}m_{b]}
 + 2\phi_{01}\,(\ell_{[a}n_{b]} - m_{[a}\mb_{b]})
 - 2\phi_{11}\,n_{[a}\mb_{b]}\right),
\end{equation} 
where
\begin{align}
\phi_{11} & = (\chi\bar\chi)^{-1}F_{ab}\ell^am^b \\
\phi_{01} & = \frac{1}{2}(\chi\bar\chi)^{-1}\left(
 F_{ab}n^a\ell^b - F_{ab}\mb^am^b\right) \\
\phi_{00} & = (\chi\bar\chi)^{-1}F_{ab}\mb^a n^b
\end{align}

Most of the equations in this paper will be given using the compacted
GHP-formalism, see [3,7,8]. In
this formalism, we concentrate on those spin coefficients of good weight, that
is, those that transform homogeneously under a spin-boost transformation of the
dyad: if
$$ \o^A\mapsto \lambda \o^A \qquad \i^A\mapsto \mu\i^A$$
a {\bf weighted} quantity $\eta$ of {\bf type} $\{r',r;t',t\}$ undergoes a 
transformation
$$  \eta \mapsto \lambda^{r'}\bar\lambda^{t'} \mu^{r}\bar\mu^{t}\eta.$$
These weights will be referred to as the {\bf Penrose-Rindler (PR)} weights, to
distinguish them from the more familiar GHP-weights $(p,q)$ for a normalised dyad
in e.g.~[3,7]. The two sets of weights are related by $p=r'-r$ and $q=t'-t$.

\section{General Equations}

The Killing equations themselves, (\ref{eq:Killing}), are unaffected by the
curvature and so are the same as in [6].
\newcounter{compkilleq}
\stepcounter{equation}
\setcounter{compkilleq}{\value{equation}}
\begin{align}
\thorn\Hl & = -\bar\kappa\Hm - \kappa\Hmb;   \tag{\thecompkilleq$a$}\\
\thorn'\Hl & =  -\bar\tau\Hm -\tau\Hmb 
     -(\phi_{01}+\bar\phi_{01})+\psi; \tag{\thecompkilleq$b$} \\
\eth\Hl &= -\bar\rho\Hm-\sigma\Hmb  +\phi_{11}; \tag{\thecompkilleq $c$}\\
\thorn\Hn &= -\tau'\Hm-\bar\tau'\Hmb 
     +(\phi_{01}+\bar\phi_{01})+\psi; \tag{\thecompkilleq $d$} \\
\thorn'\Hn &=  - \kappa'\Hm -\bar\kappa'\Hmb ;  
   \tag{\thecompkilleq$e$}\\
\eth\Hn &=  -\rho'\Hm-\bar\sigma'\Hmb    -\bar\phi_{00};
   \tag{\thecompkilleq$f$}\\
\thorn\Hm & = -\bar\tau'\Hl - \kappa\Hn  -\phi_{11};\tag{\thecompkilleq$g$}\\
\thorn'\Hm & = -\bar\kappa'\Hl - \tau\Hn +\bar\phi_{00}; 
   \tag{\thecompkilleq$h$}\\
\eth\Hm &= -\bar\sigma'\Hl -\sigma\Hn  ;   
    \tag{\thecompkilleq$i$}\\
\eth'\Hm & = -\bar\rho'\Hl-\rho\Hn 
    +(\phi_{01}-\bar\phi_{01})- \psi; \tag{\thecompkilleq$j$}
\end{align}

The spin-boost weights $(r',r,t',t)$ of the components of $\xi^a$ and $F_{ab}$  are
given in Table I (correcting a minor typo in [6]).

\begin{table}[h]
$$\vbox{
\tabskip=0pt\offinterlineskip
\halign %to \hsize
{\strut#&\vrule # \tabskip=.75em plus 2em minus 0.5em
  &\hfil $#$\hfil&\vrule# &\hfil $#$\hfil&\vrule# 
  &\hfil $#$\hfil&\vrule# &\hfil $#$\hfil&\vrule# 
  &\hfil $#$\hfil&\vrule# &\hfil $#$\hfil&\vrule# 
%  &\hfil #\hfil&\vrule# &\hfil #\hfil&\vrule# 
  &\hfil $#$\hfil&\vrule# &\hfil $#$\hfil&\vrule# \tabskip=0pt\cr
\noalign{\hrule}
\omit & height 3pt && height 3pt &&  height 3pt && %height 3pt && height 3pt &&
 height 3pt && height 3pt  && height 3pt && height 3pt && height 3pt 
&& height 3pt\cr
&&   && \Hl && \Hn && \Hm && \Hmb && \phi_{00} && \phi_{01} && \phi_{11} &\cr
\noalign{\hrule}
\omit & height 3pt && height 3pt &&  height 3pt && %height 3pt && height 3pt &&
 height 3pt && height 3pt  && height 3pt && height 3pt && height 3pt 
&& height 3pt\cr
&& r' && 0 && -1 && 1 && 0 && -1 && 0 && 1&\cr
\noalign{\hrule}
\omit & height 3pt && height 3pt &&  height 3pt && %height 3pt && height 3pt &&
 height 3pt && height 3pt  && height 3pt && height 3pt && height 3pt 
&& height 3pt\cr
&& r && -1 && 0 && -1 && 0 && 1 && 0 && -1&\cr
\noalign{\hrule}
 \omit & height 3pt && height 3pt &&  height 3pt && %height 3pt && height 3pt &&
  height 3pt && height 3pt  && height 3pt && height 3pt && height 3pt 
 && height 3pt\cr
 && t' && 0 && -1 && -1 && 0 && 0 && 0 && 0&\cr
 \noalign{\hrule}
\omit & height 3pt && height 3pt &&  height 3pt && %height 3pt && height 3pt &&
  height 3pt && height 3pt  && height 3pt && height 3pt && height 3pt 
 && height 3pt\cr
 && t && -1 && 0 && 0 && -1 && 0 && 0 && 0&\cr
 \noalign{\hrule}
}}$$
%\endinsert
\caption{weights of components}
\end{table}

The Ricci identity for $\xi^a$ implies $F_{cd;b}=R_{abcd}\xi^a$,  from which
the algebraic Bianchi identities lead  to equations for the derivatives of the
$\phi_{ij}$.  
The anti-self-dual of the above equation takes the
spinor form
\begin{equation}
\nabla_{CC'}\phi_{AB} = \left(\Psi_{ABDC}\epsilon_{D'C'} +
\Phi_{ABD'C'}\epsilon_{DC}\right)\xi^{DD'} -
\Lambda\left(\epsilon_{BC}\xi_{AC'}+\epsilon_{AC}\xi_{BC'}\right)
\label{eq:spinorintcond}
\end{equation}
Here $\Psi_{ABCD}$ is the (totally symmetric) Weyl spinor, 
$\Phi_{ABA'B'}$ the Ricci spinor and $24\Lambda=R$, the Ricci scalar (see [7]).

The components of the Weyl and Ricci spinors are given in [7] (4.11.6) and
(4.11.8) respectively, and then resolving equation (\ref{eq:spinorintcond}) we
get the (first) integrability conditions
\newcounter{intcons}
\stepcounter{equation}
\setcounter{intcons}{\value{equation}}
\begin{align}
\Psi_1\Hl-\Psi_0\Hmb +\Phi_{01}\Hl -\Phi_{00}\Hm & 
  =2\kappa\phi_{01} %+(\eps-\gamma')\phi_{11}-D\phi_{11};
  -\thorn\phi_{11};
  \tag{\theintcons$a$}\\[1pt]
\Psi_1\Hm-\Psi_0\Hn +\Phi_{02}\Hl -\Phi_{01}\Hm &  
  = 2\sigma\phi_{01} %+(\beta-\alpha')\phi_{11}-\delta\phi_{11};
 -\eth\phi_{11};
  \tag{\theintcons$b$}\\[1pt]
\Psi_2\Hl-\Psi_1\Hmb +\Phi_{01}\Hmb -\Phi_{00}\Hn +2\Pi\Hl & 
  = 2\rho\phi_{01} %(+\alpha-\beta')\phi_{11}-\delta'\phi_{11};
  -\eth'\phi_{11};
  \tag{\theintcons$c$}\\[1pt]
\Psi_2\Hm-\Psi_1\Hn +\Phi_{02}\Hmb -\Phi_{01}\Hn +2\Pi\Hm &
 = 2\tau\phi_{01} % +(\gamma-\eps')\phi_{11}-D'\phi_{11};
  -\thorn'\phi_{11};
  \tag{\theintcons$d$}\\[1pt]
\Psi_3\Hl -\Psi_2\Hmb +\Phi_{21}\Hl -\Phi_{20}\Hm -2\Pi\Hmb &
 = 2\tau'\phi_{01} %+(\gamma'-\eps)\phi_{00}-D\phi_{00};
   -\thorn\phi_{00};
\tag{\theintcons$e$}\\[1pt]
\Psi_3\Hm-\Psi_2\Hn +\Phi_{22}\Hl -\Phi_{21}\Hm -2\Pi\Hn & 
 = 2\rho'\phi_{01} %+(\alpha'-\beta)\phi_{00}-\delta\phi_{00};
   -\eth\phi_{00};
 \tag{\theintcons$f$}\\[1pt]
\Psi_4\Hl -\Psi_3\Hmb +\Phi_{21}\Hmb -\Phi_{20}\Hn & 
 = 2\sigma'\phi_{01} %(+\beta'-\alpha)\phi_{00}-\delta'\phi_{00};
  -\eth'\phi_{00};
 \tag{\theintcons$g$}\\[1pt]
\Psi_4\Hm -\Psi_3\Hn +\Phi_{22}\Hmb -\Phi_{21}\Hn &  
 = 2\kappa'\phi_{01} %+(\eps'-\gamma)\phi_{00}-D'\phi_{00};
   -\thorn'\phi_{00}
\tag{\theintcons$h$}\\[1pt]
\Psi_2\Hl-\Psi_1\Hmb +\Phi_{11}\Hl -\Phi_{10}\Hm -\Pi\Hl & 
  = \thorn\phi_{01}-\tau'\phi_{11}-\kappa\phi_{00};
  \tag{\theintcons$i$}\\[1pt]
\Psi_2\Hm-\Psi_1\Hn +\Phi_{12}\Hl -\Phi_{11}\Hm -\Pi\Hm &
 = \eth\phi_{01}-\rho'\phi_{11}-\sigma\phi_{00};
   \tag{\theintcons$j$}\\[1pt]
\Psi_3\Hl -\Psi_2\Hmb +\Phi_{11}\Hmb -\Phi_{10}\Hn +\Pi\Hmb &
 = \eth'\phi_{01}-\rho\phi_{00}-\sigma'\phi_{11};
   \tag{\theintcons$k$}\\[1pt]
\Psi_3\Hm-\Psi_2\Hn +\Phi_{12}\Hmb -\Phi_{11}\Hn +\Pi\Hn & 
 = \thorn'\phi_{01}-\tau\phi_{00}-\kappa'\phi_{11}.
 \tag{\theintcons$l$}
\end{align}
where $\Pi=\chi\bar\chi\Lambda$. These are equivalent to the equations (20)--(22) in
[5].

Note that there are four pairs of equations with the same Weyl curvature terms
($c$/$i$; $d$/$j$; $e$/$k$ and $f$/$l$).
We can eliminate the Weyl curvature terms between these pairs to give equations
equivalent to (23) in [5]:
\newcounter{maxintcons}
\stepcounter{equation}
\setcounter{maxintcons}{\value{equation}}
\begin{align}
\thorn\phi_{01} +\eth'\phi_{11} & - \kappa\phi_{00}  -2\rho \phi_{01}
-\tau'\phi_{11}\notag\\
 &= \left(\Phi_{11}-3\Lambda\right)\Hl +\Phi_{00}\Hn-\Phi_{10}\Hm-\Phi_{01}\Hmb
\tag{\themaxintcons$a$}\\[1pt]
\eth\phi_{01} +\thorn'\phi_{11} & - \sigma\phi_{00} -
  2\tau\phi_{01} - \rho'\phi_{11}\notag\\
 &=\Phi_{12}\Hl+\Phi_{01}\Hn -\left(\Phi_{11}+3\Lambda\right)\Hm -\Phi_{02}\Hmb
&\tag{\themaxintcons$b$}\\[1pt]
\eth'\phi_{01} +\thorn\phi_{00} &- \rho\phi_{00} -
  2\tau'\phi_{01} - \sigma'\phi_{11}\notag\\
 &= -\Phi_{21}\Hl-\Phi_{10}\Hn+\Phi_{20}\Hm+\left(\Phi_{11}+3\Lambda\right)\Hmb
&\tag{\themaxintcons$c$}\\[1pt]
\thorn'\phi_{01}+\eth\phi_{00}& - \tau\phi_{00} -
  2\rho'\phi_{01} -\kappa'\phi_{11}\notag\\
& =-\Phi_{22}\Hl -\left(\Phi_{11}-3\Lambda\right)\Hn+\Phi_{21}\Hm+\Phi_{12}\Hmb
\tag{\themaxintcons$d$}
\end{align}

Note that all these equations are consistent as far as spin and boost weight are
concerned, and all reduce to the equations of [6] in vacuum.

\section{Second integrability conditions}

Since a homothetic transformation preserves connection and hence curvature, we have 
${\mathcal L}_\xi R^a{}_{bcd}=0$, and resolving the spinor version of this equation
and using ($\theintcons$) to eliminate first derivatives of the $\phi_{ij}$
leads to equations I will refer to as second integrability conditions, although they
are not strictly integrability conditions in the case of a homothety. The same
equations arises from applying
the commutators to the components of the homothetic bivector of course. 
Using the Bianchi identities and the GHP-notation these equations can be reduced to
a
very compact form. Firstly, define the zero weight derivative operator
$$\GHPL_\xi= \Hn\thorn+\Hl\thorn'-\Hm\eth'-\Hmb\eth,$$
and let
$$X_{00}=\phi_{00}-\kappa'\Hl-\tau'\Hn+\sigma'\Hm+\rho'\Hmb\qquad
X_{11}=\phi_{11}+\kappa\Hn+\tau\Hl-\sigma\Hmb-\rho\Hm.$$
(Note that under the Sachs $*$ operation, $X_{11}$ and $X_{00}$ are unchanged
but $\bar{X}_{11}^*=\bar{X}_{00}$ and $\bar{X}_{00}^*=\bar{X}_{11}$).
Then we find that
\begin{align}
\GHPL_\xi \Psi_i + 2\psi\Psi_i &= iX_{00}\Psi_{i-1} -p\,\phi_{01}\Psi_i +
(i-4)X_{11}\Psi_{i+1}\label{eq:Psitwo}\\
\GHPL_\xi \Phi_{ab} + 2\psi\Phi_{ab} &= 
   aX_{00}\Phi_{(a-1)b} + b\bar X_{00}\Phi_{a(b-1)} -(p\,\phi_{01} +
q\,\bar\phi_{01})\Phi_{ab} \notag \\
&\qquad  {}+ (a-2)X_{11}\Phi_{(a+1)b} + (b-2)\bar
X_{11}\Phi_{a(b+1)}\label{eq:Phitwo} \\
\GHPL_\xi\Pi +2\psi\Pi & =0,
\end{align}
where $p=r'-r$ and $q=t'-t$ are the GHP weights
[7,8]. Note that $\Psi_i$ has PR-weight $[3-i,i-1,1,1]$ and $\Phi_{ab}$ PR-weight
$[2-a,a,2-b,b]$.
Equations~(\ref{eq:Psitwo}) are equivalent to Collinson and French's equations
($2.2$) [1] and Kolassis and Ludwig's equations (43)--(45) [4]; 
equations~(\ref{eq:Phitwo}) are equivalent to [4] equations (47)--(49). In these
references the tetrad is assumed normalised. A comparison with [4] shows that
equations~(\ref{eq:Psitwo}) are the same for a
general conformal vector in a normalised tetrad: this is as expected since these
equations actually arise
from the derivative of the Weyl tensor part of the curvature.

Note that if $\ell^a$ is a Debever-Penrose direction then $\Psi_0=0$
and~(\ref{eq:Psitwo}) implies $X_{11}=0$, or
\begin{equation}\label{eq:DP}
  \phi_{11}=-\kappa\Hn-\tau\Hl+\sigma\Hmb+\rho\Hm
\end{equation}
correcting the error in equation (11) of [6]. Similarly, if $n^a$ is a
Debever-Penrose direction then $X_{00}=0$.

\section{Type O pure radiation metrics}

Possibly the simplest non vacuum metric to consider would be that of
conformally flat pure radiation solutions.
In [2] Edgar and Ludwig performed the integration of this case, which
I here repeat with the extra assumption of the existence of a homothetic
vector, when the calculations can be pushed to completion in the sense that no
free functions remain. The main difference here is that I  use  the 
homothety to choose coordinate candidates.
  
We begin
by assuming that the dyad is normalised ($\chi=1$) and aligned to
the Ricci tensor, so $R_{ab}=\Phi_{22}\ell_a\ell_b$, leaving complete
four parameter null rotation freedom in choosing the dyad.

The Bianchi identities quickly tell us that $\kappa=\sigma=0$, and the
remaining Bianchi identities (see [7]) are 
$$\eth'\Phi_{22}=\bar\tau\Phi_{22},\qquad
\thorn\Phi_{22}=\bar\rho\Phi_{22}\qquad
\thorn\Phi_{22}=(\rho+\bar\rho)\Phi_{22}
$$ 
So $\rho=0$ and we are in Kundt's class. Now if $\tau=0$ we have plane waves,
a case which has been much studied and we will ignore. So from henceforth,
$\tau\not=0$.

Of the second integrability equations only equation (\ref{eq:Phitwo}) for
$(a,b)$
either $(1,2)$ or $(2,2)$ are non-trivial, and the first of these gives
$\phi_{11}=-\tau\Hl$.
Suppose $\phi_{11}$, and hence $\Hl$, vanishes. Then equation
(\theintcons$d$)
and $\tau\not=0$ implies $\phi_{01}$ vanishes and then by (\theintcons$l$),
$\phi_{00}=0$. But now (\theintcons$d$) gives $\Hmb=0$, and the Killing equation
(\thecompkilleq$h$) means that $\Hn=0$ and the homothetic vector vanishes.

Thus $\Hl$ is not identically zero: no symmetry vector is
orthogonal to $\ell^a$. This means $\phi_{11}\not=0$ and we can perform
a proper null rotation about $\ell^a$ to set $\phi_{01}$ to be identically
zero: we will be left with boost and rotation freedom, which is what we
would want for a GHP integration procedure.

With $\phi_{01}=0$, the integrability equations involving the derivatives of
$\phi_{01}$, (\theintcons$i$) -- (\theintcons$l$) show that
$\sigma'=\rho'=\tau'=0$ and $\phi_{00}=\kappa'\Hl$. 

The remaining integrability conditions (\theintcons) and (\ref{eq:Phitwo}) allow
us to find all directional derivatives of all the remaining scalars. We have
\begin{alignat*}{5}
 \thorn\tau &=0  & \qquad\quad \thorn\kappa' &=0  &\qquad\quad \thorn\Phi_{22} &
= 0 \\
\eth\tau &=\tau^2  &\qquad\quad \eth \kappa' &=\tau\kappa'-\Phi_{22}  &\eth
\Phi_{22} & = \tau\Phi_{22} \\
\eth'\tau &=\tau\bar\tau  & \eth' \kappa' &=\kappa'\bar\tau  &\eth' \Phi_{22} &
= \bar\tau\Phi_{22} \\
\Hl\thorn'\tau &=\tau (W-\psi)  & \qquad \quad
\Hl\thorn'\kappa' &=-\Hmb\Phi_{22} +\kappa'(W-\psi)
&\qquad 
\Hl\thorn'\Phi_{22} & = \Phi_{22}(W-2\psi)
\end{alignat*}
where $W=\bar\tau\Hm+\tau\Hmb$ has weight $(0,0)$. The homothetic equations
reduce to
\begin{alignat*}{5}
  \thorn\Hl&=0 &\qquad\quad \thorn\Hn & = \psi&\qquad\quad \thorn\Hm& =\tau\Hl  
\\
\eth\Hl&=-\tau\Hl &\qquad\quad \eth\Hn & = -\bar\kappa'\Hl&\qquad\quad \eth\Hm&
=0  \\
\eth'\Hl&=-\bar\tau\Hl&\qquad\quad \eth'\Hn & =-\kappa'\Hl &\qquad\quad
\eth'\Hm& = -\psi \\
\thorn'\Hl&=-W+\psi &\qquad\quad \thorn'\Hn & =-\kappa'\Hm-\bar\kappa'\Hmb
&\qquad\quad \thorn'\Hm& = -\tau\Hn
\end{alignat*}
In [2], Edgar and Ludwig 
used the tetrad freedom to
set
$\tau'=\sigma'=\rho'=0$, $\Phi_{22}-\tau\kappa'-\bar\tau\bar\kappa'=0$ and
$\thorn'(\tau/\bar\tau)=0$ as a preparation to performing the integration.
In our approach, we have obtained $\tau'=\sigma'=\rho'=0$ without the need to
solve a system of differential equations, and it is not difficult to check that
we also have $\thorn'(\tau/\bar\tau)=0$.
 As for Edgar and Ludwig's other term, we note that
$\Phi_{22}-\tau\kappa'-\bar\tau\bar\kappa'$ is of weight $(-2,-2)$. Define the
real scalar $Z=\Hl^2(\Phi_{22}-\tau\kappa'-\bar\tau\bar\kappa')$ of weight
$(0,0)$.
Then we can easily check that $Z$ is annihilated by all the derivative
operators and is hence constant.

We will not attempt to show $Z$ is zero, as we wish to pick a set of coordinate
candidates (real weight $(0,0)$ scalars) more attuned to the homothetic or Killing
vector than those of Edgar and Ludwig.

Firstly, as in [2], define the convenient scalars $P
=\sqrt{\tau/2\bar\tau}$, complex of weight $(1,-1)$ and 
$A=(2\tau\bar\tau)^{-1/2}$, real of weight $(0,0)$. The scalar $P$ has the happy
property of being annhilated by all the GHP operators, whereas 
\begin{equation}
  \thorn A =0 ,\qquad \thorn' A = \frac1{\Hl}(A\psi -u),\qquad 
\eth A=  -P.\label{eq:QDEss}
\end{equation}
Our coordinate candidates are $w$, $x$ and $y$ where 
$$ w = \Re\left(\frac{\Hm}{P}\right),\qquad 
z=x+iy = -2\frac{\kappa'P}{\Phi_{22}},$$
and also a real scalar $u$ of weight $(0,0)$ satisfying the equations (cf.~[2])
$$\thorn u=0,\qquad \eth u=0,\qquad \thorn'u = 1/\Hl.$$
These latter equations are consistant, as can be checked by verifying
the commutators are satisfied. The commutators acting 
on weight $(0,0)$ scalars here simplify to
$$\left[\thorn,\thorn'\right] = A^{-1}\left(
\frac{1}{\bar P}\eth+\frac{1}{P}\eth'\right),\qquad
\left[\thorn,\eth\right] = %\left[\thorn,\eth'\right] =
\left[\eth,\eth'\right]=0,\qquad
\left[\thorn',\eth\right]=-\frac{1}{AP}\thorn'-\kappa'\thorn$$
and their conjugates.

Our table of deriviatives reads
\begin{alignat*}{5}
\thorn u & =  0 & \qquad
\thorn'u & = 1/\Hl & \qquad
\eth u & = 0 \\ 
\thorn w & = \Hl/A & \qquad
\thorn'w & = -B/\Hl & \qquad
\eth w & = -\psi P \\
\thorn x & = 0 & \qquad
\thorn' x & = \frac{1}{\Hl}(\psi x+w) & \qquad
\eth x& = P\\
\thorn y & = 0 & \qquad
\thorn'y & = -\frac{1}{\Hl}(\psi y -v) & \qquad
\eth y & = -i P
\end{alignat*}
where $B=\Hn\Hl/A$ and $v=\Im(\Hm/P)$ are both real of weight $(0,0)$.
With the coordinates in the order $(u,w,x,y)$, the tetrad is thus
\begin{align*}
  \ell^a&=\Hl\left(0,A^{-1},0,0\right)\qquad
 n^a = {\Hl}^{-1}\left(1,-B,\psi x+w,\psi y-v,\right)\qquad
m^a  = P\left(0,-\psi,1,-i\right)
\end{align*}
The $\Hl$ and $P$ terms are a manifestation of the residual boost and spin freedom
of course.

The other homothetic vector components are $\Hn=AB/\Hl$ and $\Hm=(w+iv)P$,
and the homothetic vector simplifies to
$$ \xi^a = \left(1,\psi w,\psi x,\psi y\right)$$
so that in the Killing case $u$ is a cyclic coordinate. 

 To complete the integration we need to find $A$, $B$ and $v$.
Since they are all weight (0,0) this is straightforward: 
$$ v=\psi y -c_3 e^{\psi u},\qquad
%k = c_1e^{-\psi t},\qquad 
A=c_2e^{\psi u}-x,\qquad
 B = \psi w +\psi^2 x +(x^2+y^2)c_1e^{-\psi u}+c_4e^{\psi u},$$
where the $c_i$ are constant.
Note that in the Killing case, $v$ is constant.

The metric is then
$$ ds^2 = H\,du^2 +2A\,du\,dw+ 2(w+c_2\psi e^{\psi u})du\,dx+ 
2c_3e^{\psi u}du\,dy - dx^2-dy^2
$$
where 
$$
  H  = 2A\left(c_1(x^2+y^2)e^{-\psi u}+c_4e^{\psi u} )\right) - 
c_3^2e^{2\psi u}  -w^2 -\psi^2x^2-2\psi w x.
$$

\subsection{Larger algebra}

This leaves us to consider the solutions with further homothetic or
Killing vectors. Since the bracket of two homothetic vectors is Killing, we
begin by assuming the first symmetry vector is Killing, and then look for a second
(possibly homothetic) vector $\xi^a$.
So
the metric is 
$$ ds^2 = H\,du^2 +2(c_2-x)\,du\,dw+ 2w\,du\,dx+ 
2c_3du\,dy - dx^2-dy^2$$
where 
$$
  H  = 2\left((x^2+y^2)c_2+c_4\right)(c_2-x) -
 w^2 - c_3^2 %,\qquad A=c_2-x
$$
and the Killing vector is $K^a=\partial_u$. For the tetrad we perform a
rotation to make $\tau$ (and hence $P$) real and also a boost to allow us
to integrate more easily: so
$$
 \ell^a=\left(0,1,0,0\right)\qquad
 n^a = \frac{1}{c_2-x}\left(1,-c_4-c_1(x^2+y^2),w,c_3\right)\qquad
m^a  = \frac1{\sqrt{2}}\left(0,0,1,-i\right).
$$
The advantage of this tetrad is that the
improperly weighted spin coefficients $\varepsilon$, $\alpha$, $\beta$
and $\gamma$ all vanish, leaving only
$$ \tau = \frac1{\sqrt{2}(c_2-x)}\quad \text{and}\quad
\kappa'=-\sqrt{2}c_1\,\frac{x+iy}{c_2-x}.
$$
So the four GHP operators reduce to the basic Newman-Penrose operators $D$, $\delta$,
$\delta'$
and $D'$ (see [7,8]) for all scalars. Furthermore, the same argument as used in
the previous section tells us that $\Hl$ is non-zero and $\phi_{11}=-\tau\Hl$.

Integrating (\thecompkilleq$a$) and
(\thecompkilleq$b$) and using the reality of $\Hl$ gives $\Hl=F_1(u)(c_2-x)$ for
some $F_1(u)$.
Next we integrate (\thecompkilleq$g$), (\thecompkilleq$i$) and
(\thecompkilleq$j$) to get
$$\Hm = \frac1{\sqrt{2}}\left(F_1(u)w- \psi(x-iy)\right) +F_2(u)$$
for some complex $F_2(u)$. However, (\theintcons$d$) and the Killing equations
imply that $F_2(u)=\frac1{\sqrt{2}}\,c_2\psi+i b(u)$ for real $b$. 
And now equation (\ref{eq:Phitwo}) for $(a,b)=(2,2)$ gives
$\phi_{01}=\frac14\psi$ and so (\theintcons$l$) gives $\phi_{00}=\kappa'\Hl$.

The remaining equation for $\Hl$, (\thecompkilleq$b$), shows that 
$F_1=a_1-\frac12\psi u$ where $a_1$ is a
constant, so $\Hl=\xi^a\ell_a=(a_1-\frac12\psi u)(c_2-x)$. But
$K^a\ell_a=c_2-x$, which tells us that we cannot have a second
Killing vector ($\psi=0$), as then a linear combination of $\xi^a$ and $K^a$
would be orthogonal to $\ell^a$, which we saw is not possible. Hence 

\begin{theorem}
  Type O pure radiation metrics can admit at most a 2
parameter group of homothetic motions, and if the dimension is 2 it is a proper
homothetic group.
\end{theorem}

We also see that for the second symmetry vector (the proper homothety), we can
take $$\Hl=-\frac12\psi u(c_2-x).$$
The imginary part of (\thecompkilleq$h$) can now be solved 
for $b_2(u)$ to give
$b_2(u) = -\frac1{\sqrt{2}}\psi c_3 u + a_2$ for constant $a_2$.
The real part of (\thecompkilleq$h$) now gives us 
$$\Hn = -\frac12\psi\left( c_1(x^2+y^2)u+c_4u-w\right).$$ 
The only remaining Killing or integrability equation is (\thecompkilleq$e$) for
$D'\Hn$, and this tells us that
$c_2=c_3=c_4=a_2=0$ and thus
$$\xi^a = \frac12\psi\left(-u,3w,x,y\right).$$
The remaining constant $c_1$ is non-zero (or the metric is flat) and can be
absorbed in the coordinates. 
The type O pure radiation solution with the largest possible homothetic
symmetry group is thus 
$$ ds^2 =-\left(2x(x^2+y^2)+w^2\right)\,du^2 -2x\,du\,dw+ 2w\,du\,dx
 - dx^2-dy^2$$
with a non-abelian homothetic algebra
generated by
$\{\partial_u, -u\partial_u+3w\partial_w+x\partial_x+y\partial_y\}$.

\section{Acknowledgements}

The author is indebted to John Carminati of Deakin Universtity for the idea
of the second integrability equations given here.
Calculations were carried out using Maple, and in particular the GHPII package
of Vu and Carminati [9]. Maple is a registered trademark of Waterloo Maple Inc.

\section{References}

\begin{enumerate}
\item[{[1]}] Collinson CD and French DC {\it J.~Math.~Phys.} {\bf 8} (1967)
p.~701
\item[{[2]}] Edgar SB and Ludwig G {\it Gen.~Rel.~Grav.} {\bf 29} (1997) 
p.~1309
\item[{[3]}] Geroch R, Held A and Penrose R {\it J.~Math.~Phys.} {\bf 14}  (1973)
p.~874
\item[{[4]}] Kollasis C and Ludwig G {\it Gen.~Rel.~Grav.} {\bf 25} (1993) p.~625
\item[{[5]}] Ludwig G {\it Class.~Quantum Grav.} {\bf 19} (2002) p.~3799
\item[{[6]}] Steele JD 2002 {\it Class.~Quantum Grav.} {\bf 19} p.~259
\item[{[7]}] Penrose R and Rindler W 1984 {\it Spinors and Space-Time} vol 1
(Cambridge: Cambridge University Press)
\item[{[8]}] Stephani H, Kramer D, MacCallum M, Hoenselaers C and Herlt E
2003
{\it Exact Solutions of Einstein's Field Equations, 2nd Edition} (Cambridge:
Cambridge University Press)
\item[{[9]}] Vu K and Carminati J {\it Gen.~Rel.~Grav.} {\bf 35} (2003) p.~263
\end{enumerate} 

\end{document}